\documentclass[twocolumn,english,floatfix,aps,prb]{revtex4}
\usepackage{graphicx}
\usepackage{mathrsfs}
\usepackage{natbib}
\setcitestyle{square,numbers,sort&compress}
\usepackage{amsmath}
\usepackage{amssymb}
\usepackage{color}
\makeatletter

\usepackage{babel}
\usepackage{algorithm}
\usepackage{algorithmic}

\newcommand\figref[1]{Fig.~\ref{#1}}

\newcommand\dcp[1]{${\mathcal {#1}}$}
\newcommand\dcpr[1]{${\mathcal {#1}^r}$}
\newcommand\dcph[1]{${\mathcal {#1}}_{\langle\rangle}$}
\newcommand\dcpi[1]{${\mathcal {#1}}_{\left[\,\right]}$}
\newcommand\dcphr[1]{${\mathcal {#1}}_{\langle\rangle}^r$}
\newcommand\dcpir[1]{${\mathcal {#1}}_{\left[\,\right]}^r$}

\makeatother
\begin{document}

\title{Environment Overwhelms both Nature and Nurture in a Model Spin Glass}

\author{Jie Yang, A. Alan Middleton}

\affiliation{Department of Physics, Syracuse University, Syracuse, New York 13244, USA}

\begin{abstract}
The microscopic dependence of glassy equilibration on sample history, both the initial configuration and the specific sequence of random noise, is examined. The temporal evolution of spin configurations in a two-dimensional Ising spin 
glass is simulated using patchwork dynamics, a coarse-grained heuristic for glassy dynamics, allowing simulations over a wide range of length scales.
Most of the nearest-neighbor spin correlations are independent of the details of evolution, due to the formation of rigid domains. The correlations on a fractal set of domain walls are found to be variable and to depend distinctly on the noise history and the initial state. Correlations in samples with independent initial configurations are found to persistently differ when subject to identical noise histories, for coarsening scales of up to hundreds of lattice units. However, samples with identical initial configurations and subject to independent noise histories have correlations that converge to each other as a power law in scale. The initial ``nature'' of the state is retained in the domain walls during coarsening under distinct noises (``nurture''), while the location of the domain walls at each scale is determined by the frozen disorder (``environment''), independent of initial state or noise history. We also provide evidence that coarsening with local dynamics at finite temperature gives persistence of overlaps roughly consistent with coarsening at zero temperature.
\end{abstract}
\maketitle

The behaviors of spin glasses are both intricate and challenging to study, due to the simultaneous presence of quenched disorder and competing ferromagnetic and antiferromagnetic interactions \cite{Edwards1975}. Experimental
research on spin glass materials, metallic alloys with magnetic impurities dispersed randomly in the main non-magnetic component \cite{Anderson1970,Binder1986}, indicate extremely slow dynamics, suggesting that direct numerical simulation using local spin flips will take very long times to equilibrate. A rich set of nonequilibrium effects, including aging, rejuvenation, and memory, are observed \cite{Vincen2007}, suggesting complex, temperature-dependent, hierarchies of spin-glass configurations. Additionally, phenomenological and proposed analytic explanations, including 
the replica symmetry breaking description \cite{Parisi1979, Mezaed1984} and the droplet picture \cite{Bray1984, Fisher1988}, have unresolved differences in predictions, for
example with respect to the existence of a phase transition at finite magnetic field \cite{Almedia1978}. For these reasons, it is important to understand the details of
nonequilibrium evolution, including the dependencies on initial microscopic configurations and the thermal noise, in order to better understand how the history of prior conditions is stored in spin glasses.

Computer simulations, mainly using Glauber dynamics, have been frequently applied and have provided insight into both equilibrium and non-equilibrium properties of spin glasses by testing theoretical pictures. But numerics are hindered by glassy numerical dynamics and
the fact that computing ground states of spin glasses is, in the general case, an NP-hard problem \cite{Papadimitriou1982, Barahona1982}, which takes a time exponential in the number of degrees of freedom to solve.
An exceptional case is the two-dimensional Ising spin glass (2DISG), where computations for large samples is computationally tractable: well-developed numerical algorithms exist \cite{Barahona1982,Kasteleyn1963, Thomas2007} that, in time polynomial in
the number of spins, efficiently compute ground states at zero-temperature and, at finite temperature, compute partition functions and sample configurations. More significantly, the 2DISG and even one-dimensional glassy 
system \cite{Zou2010} exhibit classic glassy behavior, such as aging and memory effects \cite{Thomas2008}, in the low-temperature limit. The 2DISG provides a simplified but effective model for exploring spin glasses. In this paper, we are interested in examining how the initial spin configurations (``nature") and the 
noise histories (``nurture") each determine the nonequilibrium configurations of the 2DISG as the system evolves over time. For example, when a given random initial state is chosen for a system with given disorder, independent dynamical noise histories can be applied to multiple copies of the system. Alternatively, for a chosen noise history, the 2DISG initialized using multiple 
uncorrelated random states can be evolved under a single noise history. In both ensembles, coarsening of phase domains takes place, so that most nearest-neighbor spins have a single relative orientation, in both sets, at a given coarsening scale. However, we find that these correlations vary along fractal domain walls that separate the large fixed regions depend on the sample history. We explore how the initial states, the noise histories and the intrinsic disorder (``environment") determine 
the variability during evolution of these domain walls in the 2DISG.  

We use the Edwards-Anderson Ising spin glass model \cite{Edwards1975}, with Hamiltonian $\mathcal {H}=-\sum_{\langle i,j\rangle}J_{ij}s_is_j$, where the $s_i$
are spin variables with $s_i=\pm1$ and the $J_{ij}$, the frozen disorder (``environment'') are nearest neighbor couplings chosen from a mean-zero Gaussian distribution. The
2DISG we simulate has $N=L\times L$ spins arranged on a square lattice with toroidal boundary conditions. The 2DISG is paramagnetic for all finite temperatures, though for a
range of temperatures near zero temperature, where the spin glass correlation length exceeds $L$, it exhibits a rigidity and temperature chaos \cite{Thomas2011} expected in higher dimensional models. The relaxation of the 2DISG at such low temperatures has previously been studied by patchwork dynamics, an efficient heuristic argued to be effective for temporal 
evolution \cite{Thomas2008,Yang2017}. Coarse-grained low-energy spin configurations of the 2DISG are found by exactly optimizing randomly chosen subsystems --- patches --- at a selected spatial scale, covering the system multiple times. By successive rounds, increasing the scale of the patches in each round, growth of spin-glass domains is generated.
For this paper, in order to investigate the influence of
an initial state (``nature'') on the long-time nonequilibrium configurations, independent sequences of patches were applied to multiple copies of the same initial state and then
the statistics of 
noise-history-averaged long-time behavior are quantified; to understand the influence of a noise history (``nurture'') on microscopic configurations, multiple independently chosen initial states were subjected to the same noise history (sequence of patches) and likewise the initial-state-averaged statistics were computed. In the following, we denote the noise-history-averaged statistics 
by angular brackets {$\langle\ldots\rangle$, the initial-state-averaged statistics by square brackets $\left[\ldots\right]$, and the quenched-disorder average by overlines $\overline{\ldots}$.

Before investigating in detail the influence of nature and nurture on the spin correlations at domain walls, we first carried out a heritability 
study \cite{Ye2013, Ye2017} for spin overlaps and a study of correlation functions as a measure of domain growth.
Consider a given group of identical 2DISG systems, ``clones'', initialized by the same random configuration. The clone systems undergo 
independent patchwork dynamics at zero temperature. We aim to measure the decay of similarity between the clones subject to independent noise histories, both in their bulk overlap and the size of similarly correlated regions. For the first measure, heritability, we use the sample-averaged spin overlap $q(\ell)=N^{-1}\overline{\sum_i s_i^1 s_i^2}$ between two clone systems as a function of patch scales $\ell$, where $s_i^1$ 
and $s_i^2$ denote the states of $i^{th}$ spin in clone $1$ and clone $2$ respectively. For the second measure, we compute the sample-averaged spin-glass correlation function 
$G(r)=N^{-1}\overline{\sum_i{ \left( M^{-1}\sum_c s_i^c s_{ir}^c \right)^2}}$, where $s_i^c$ and $s_{ir}^c$ are the states of spins at spacing $r$ in the $c^{th}$
clone and $M$ denotes the number of independent clones. For this correlation, we choose spins separated horizontally and vertically by distance $r$. The spin correlation function $G(r)$ can be written as 
$G(r)=M^{-1}+\left(1-M^{-1}\right)\overline{s_i^cs_i^{c^{\prime}}s_{ir}^cs_{ir}^{c^{\prime}}}$, where $c$ and $c^{\prime}$ denote
indexes for different clones. To eliminate the dependence of $G(r)$ on the number of clones $M$, we use the unbiased statistic 
$G^{\prime}(r)$, $G^{\prime}(r)=\overline{s_i^cs_i^{c^{\prime}}s_{ir}^cs_{ir}^{c^{\prime}}}=\left(M-1\right)^{-1}\left(MG(r)-1\right)$. 

\begin{figure}
	\centering
	\includegraphics[width=\columnwidth]{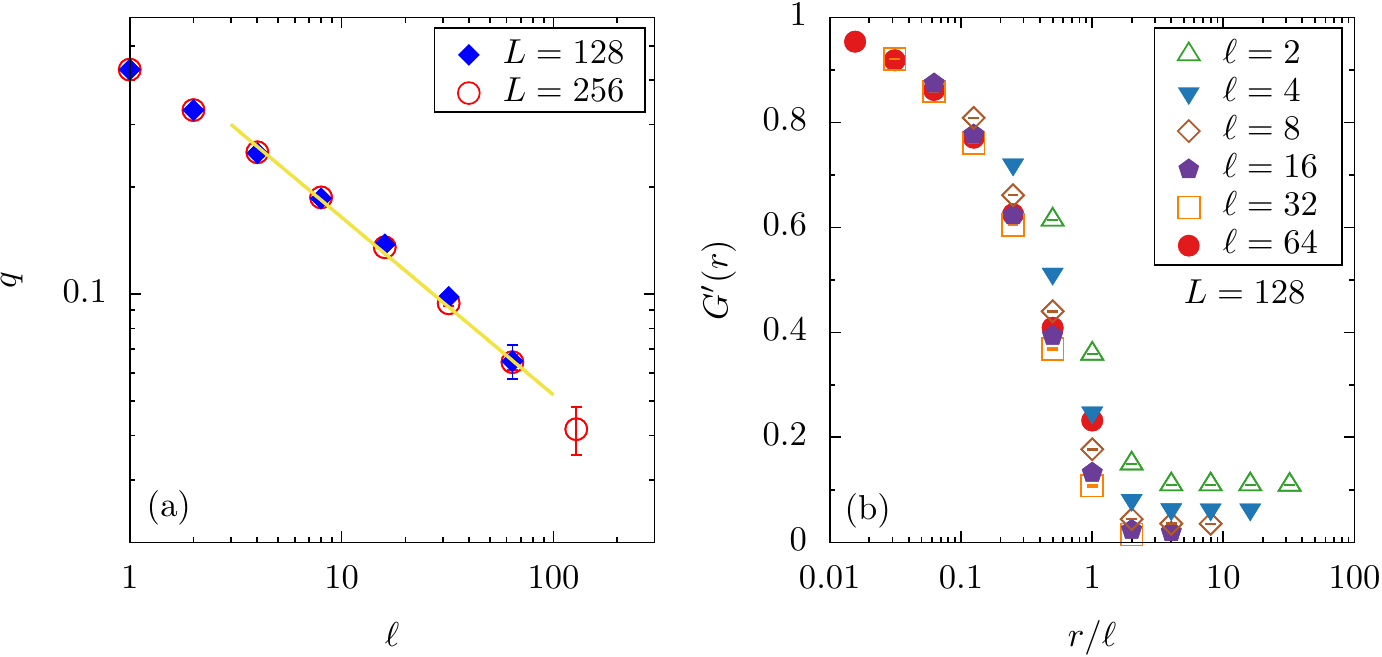}
	\caption{(Color online). 
		Plots of the heritability of the spin overlap $q$ and the spin correlation function $G^{\prime}(r)$. 
		(a) Two spin glass instances, with identical couplings and initial spin configurations, evolve under independent noise histories, i.e., sequences of optimizing patches at zero temperature, where the patch size $\ell$ is doubled on each sweep. The spin overlap $q$ between the systems, for  sizes $L=128, 256$, averaged over $5000$ pairs, decays approximately as a power law, $q(\ell)\sim{\ell}^{-\theta_h}$ with a fitted heritability exponent $\theta_h \simeq 0.5$.
		(b) Sets of spin glass systems, each set with $M=160$ clones with the same $J_{ij}$ and the same random spin configuration, are then updated under independent noise histories. The calibrated sample-averaged spin correlation function 
		$G^{\prime}(r)$ for $\ell=2,4,8,16,32,64$, averaged over 100 sets of clones, is plotted versus $r/\ell$. For spin spacing $r<\ell$, the spin correlation function $G^{\prime}(r)$ appears to converge to a fixed value at each $r/\ell$. For $r>\ell$, the value of $G^{\prime}(r)$ converges to an $\ell$-dependent plateau, consistent with $G^{\prime}(r>\ell) \approx q^2(\ell)$.
	\label{fig:q_G}
	}
\end{figure} 

We implement the heritability study of spin overlap in the 2DISG with system sizes $L=128,256$ and with $M=160$ independent runs of patchwork dynamics. The computed dependence of $q(\ell)$ on $\ell$ and $G^\prime(r)$ on $r/\ell$
are plotted in \figref{fig:q_G}. We find that $q(\ell)$ is approximately described by a power law, $q\sim{\ell}^{-\theta_h}$ for patch scales 
$\ell=4$ to $\ell=L/4$, yielding a heritability exponent $\theta_h=0.5\pm0.05$. As a check on consistency with other work, we note that our computed spin overlap $q\left(\ell=1\right)\simeq0.43$ is in accord with the result obtained by Glauber dynamics at zero temperature \cite{Ye2017}; patch dynamics at scale $\ell=1$ is the same dynamics as Glauber dynamics.
The plot of spin-correlation $G^{\prime}(r)$, for $L=128$ and  sequential patchwork dynamics length scales $\ell=2,4,8,16,32,64$, indicates convergence to the scaling form $G'(r;\ell)={\cal G}(r/\ell)$ for $r < \ell$. For $r/\ell\gg1$, the spin correlation appears to reach a constant plateau value that depends on $\ell$. 
We find that the value of 
$G^{\prime}(r)$ on the plateau is consistent with $G^{\prime}(r) = q^2(\ell)$. This consistency resonates with our expectation that spins separated by distances $r\gg\ell$ have not become correlated by the dynamics, so correlations between clones are those remaining from the identical initial conditions. In this limit, $G'(r)$ factorizes,
$G^{\prime}(r,\ell)\approx\overline{s_i^cs_i^{c^{\prime}}}\cdot\overline{s_{ir}^cs_{ir}^{c^{\prime}}}= q^2(\ell)$.  

We conjecture that the heritability is connected with the phenomena of persistence \cite{Derrdia1994}. Persistence describes 
the probability that a spin has never flipped from its initial value. For patch dynamics, the probability is found to decay as a power law in the patch size (coarsening scale) and the persistence 
exponent is numerically fitted by $\theta_p\simeq0.5$ \cite{Thomas2008}. We find that the heritability exponent is very close to the persistence exponents in 2DISG. The 
phenomena that the heritability  exponent $\theta_h$ approximates the persistence exponent $\theta_p$ is also seen in the two-dimensional Ising ferromagnet \cite{Ye2013}, where the 
evolution of the two-dimensional Ising ferromagnet follows Glauber dynamics at zero temperature. 
Further work is required to explore the specific relation between heritability and persistence.

To further develop the picture of equilibration and history dependence in this glassy system, we examined the relative importance of initial states, noise histories, and intrinsic disorder in determining the long-time states of the 2DISG, as measured by the variability of nearest-neighbor correlations.
Consider the ensemble average of the square of the nearest-neighbor correlation $\mathcal G=\overline{\|s_is_j\Vert^2}$, where $s_i$ and $s_j$ denote the orientations a pair of nearest-neighbor spins and $\|\ldots\Vert$ indicates a choice of either
average we consider, the average over noise history with fixed initial state, $\langle\ldots\rangle$, or the average over initial states with fixed noise history, $\left[\ldots\right]$. The overline indicates an average over disorder. The 2DISG with continuous disorder has two degenerate 
states related by global spin-reversal symmetry (this holds in any finite system and has strong numerical support \cite{Middleton1999} for infinite systems, though this is not yet proven mathematically \cite{Arguin2016}), so that spins in the ground state have $\mathcal G\left(\ell\rightarrow\infty\right)=1$. During equilibration, at finite $\ell$, however, the nearest-neighbor spin correlations separating large domains are not fixed and we study those correlations to address equilibration and history dependence.

To describe our protocols for visualizing and quantifying the dependence of nonequilibrium spin configurations on initial state and noise history, we denote the dependence of a non-equilibrium spin configuration on bond disorder $J_{ij}$, noise history $\eta$, and initial configuration $\sigma$ given by a spin configuration $s_i(\ell=0)$, using the notation $s_i(J,\eta,\sigma)$.
The nearest-neighbor spin correlations between neighboring pairs of spins $s_i$ and $s_j$, for a given disorder $J$ and initial state $\sigma$, averaged over $n_\eta$ noise histories is then
\begin{equation}
\langle s_i s_j \rangle(J,\sigma) = n_\eta^{-1}\sum_{m=1}^{n_\eta} s_{i}(J, \eta_m,\sigma)s_{j}(J, \eta_m,\sigma)\ .
\end{equation}
An example of the dependence of this local noise averaged configuration on initial spin configurations and coarsening scale $\ell$ is displayed in \figref{fig:nbCorI_v}, where the square of the correlation, $\langle s_i s_j\rangle^2$, is displayed for two
uncorrelated initial spin configurations $\sigma_1$ and $\sigma_2$.The noise-history-averaged nearest-neighbor correlation configurations were computed for a single disorder and $L=64$, with an average over $n_\eta = 10^3$ histories. The visualization of $\langle s_i s_j\rangle^2$ is shown by dual bonds that cross the nearest-neighbor bonds $(i,j)$, with the gray scale shading determined by the intensity 
of the correlations $\langle s_is_j\rangle^2$, such that bonds relatively independent of history are light colored for each initial configuration and dark for bonds that are dependent on history. The bonds with $\langle s_is_j\rangle^2=1$, to within our sampling error, are rigid.
As patch scale $\ell$ increases, larger light regions, domains of fixed relative spin orientations, are evident, connected by (darker) bonds that vary with history. That is, rigid clusters develop, with the clusters independent of the initial spins and clusters separated by statistically ``flexible'' bonds. As $\ell$ increases, the locations and the shading (indicating degree of statistical flexibility) of the domain walls appear to be nearly independent of initial configuration $\sigma$, when averaging over noise histories.

\begin{figure}
	\centering
	\includegraphics[width=\columnwidth]{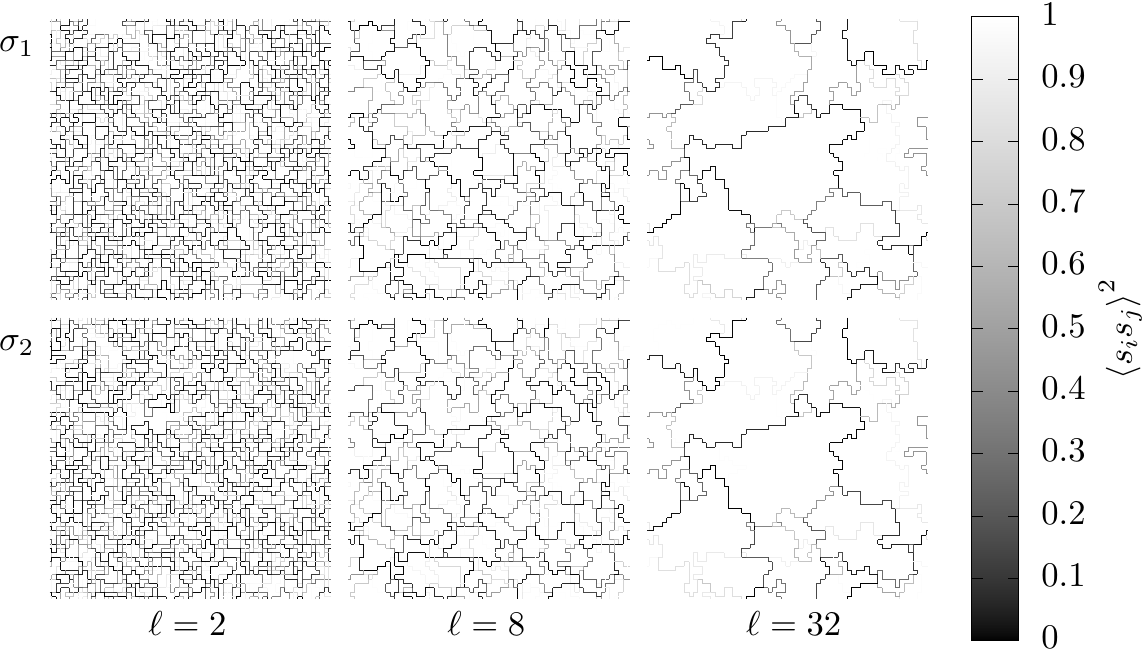}
	\caption{(Color online).
		Evolution of noise-history-averaged nearest-neighbor correlation $\langle s_i s_j\rangle^2$ configurations in patchwork dynamics for a typical 
		2DISG with system size $L=64$ and two initial configurations $\sigma_1$ and $\sigma_2$, displayed in two rows, each initial configuration evolved using
       $n_\eta=10^3$ independent noise histories. The value of $\langle s_i s_j\rangle^2$ is visualized by a dual bond that crosses the bond between nearest-neighbor spins, with the gray scale of dual bonds corresponding to 
        $\langle s_i s_j\rangle^2$; light regions are domains with spins having fixed relative orientation (to within our statistics) and darker regions indicate history-dependent domain walls.  As patch scale $\ell$ grows, 
        the upper and lower $\langle s_i s_j\rangle^2$ configurations gradually lose the differences given by the initial spin states $\sigma_1$ and $\sigma_2$, with converging correlations at each bond. 
	}
	\label{fig:nbCorI_v}
\end{figure}

\begin{figure}
	\centering
	\includegraphics[width=\columnwidth]{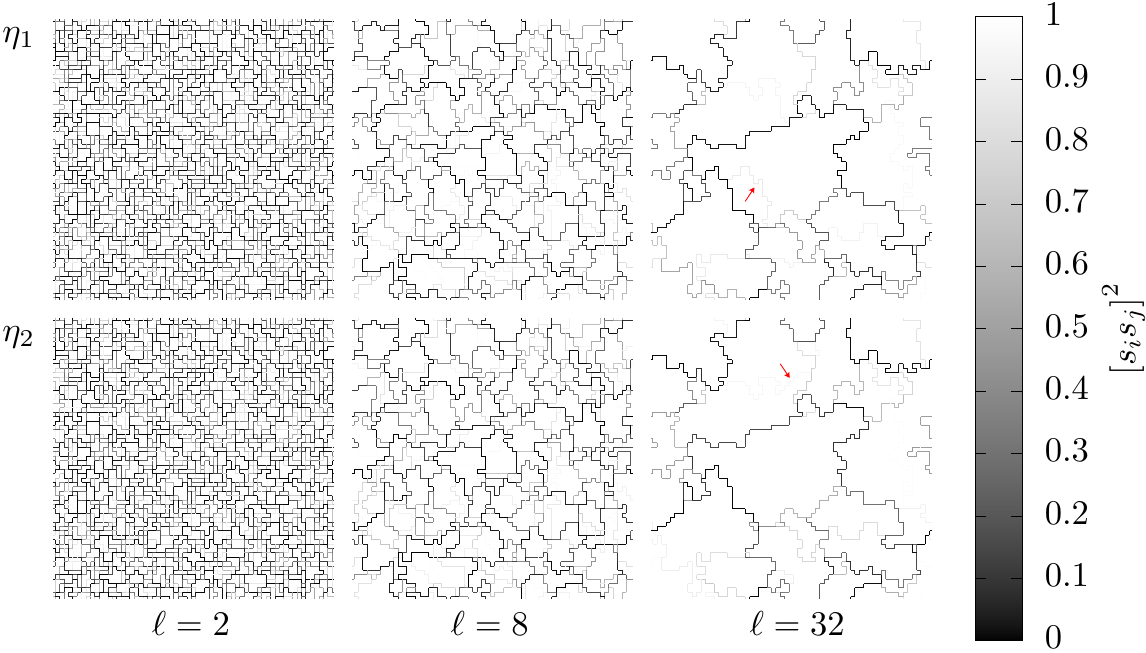}
	\caption{(Color online).
		Snapshots of initial-state-averaged nearest-neighbor spin correlation $\left[s_i s_j\right]^2$ configurations for the 2DISG with system size $L=64$, for a selected disorder,
       for two noise histories, $\eta_1$ and $\eta_2$, with increasing patch scale $\ell$. Each chosen noise history is applied to $10^3$ uncorrelated random initial spin states. The nearest-spin correlations $\left[s_i s_j\right]^2$ are visualized using the scheme used in \figref{fig:nbCorI_v}. Again, growing rigid domains, independent of noise or initial configuration, are seen, separated by variable domain walls. The $\left[s_is_j\right]^2$ configurations experiencing the noise histories $\eta_1$ and 
        $\eta_2$ maintain observable differences as patch scale $\ell$ increases and even approaches the system size. The red arrows on the plot indicate sample domain wall differences.
        	}
	\label{fig:nbCorP_v}
\end{figure}

A comparable visualization of the square of the initial-state-averaged nearest-neighbor correlation, with fixed disorder $J$ and fixed noise history $\eta$,
\begin{equation}
\left[s_is_j\right](J,\eta) = n_\sigma^{-1}\sum_{m=1}^{n_\sigma} s_{i}(J, \eta,\sigma_m)s_{j}(J, \eta,\sigma_m)
\end{equation}
is shown in \figref{fig:nbCorP_v}. One set of $n_\sigma=10^3$ random initial states is nurtured using the noise history $\eta_1 $ and another set of the same size by the noise history $\eta_2$, with each row corresponding to the evolution for the two sets, given the same disorder $J_{ij}$ for each set. We observe that the upper and lower $\left[s_is_j\right]^2$ configurations reach similar sets of domain walls with each noise history, with the domain walls separating nearly identical rigid domains. However, the noise-history-dependent $\left[s_is_j\right]^2$ values indicate remnant
differences in the initial-configuration-averaged domain wall statistics that result from distinct noise histories.

To more precisely evaluate how 2DISG systems whose initial nature is given by distinct random spin states or nurtured by distinct noise histories either converge or maintain discrepancies, we define pair-wise discrepancy statistics to compare directly the nearest-neighbor correlation
configurations $\langle s_is_j\rangle^2$ and $\left[s_is_j\right]^2$.
The noise-history-averaged \dcph{D} and initial-state-averaged \dcpi{D} are
computed by
\begin{equation}
    {\mathcal D}_{\langle\rangle}=
\overline{
  \left\{ \langle s_i s_j\rangle (J,\sigma_1) - \langle s_i s_j \rangle(J,\sigma_2)\right\}^2/2}
\end{equation}
and 
\begin{equation}
    {\mathcal D}_{\left[\,\right]}=\overline{\left\{\left[s_is_j\right](J,\eta_1)-\left[s_is_j\right](J,\eta_2)\right\}^2/2}\ ,
\end{equation}
respectively.
In both cases, distinct pairs of initial conditions or noise histories are used for distinct samples and there is an implied average over nearest neighbor pairs $(i,j)$.
These statistics are dominated by the contributions of the fluctuating walls between domains, though these contributions are reduced on average by the low density of the walls at large $\ell$.
To separate out the contributions of nonrigid correlations to the discrepancy statistic, we also compute
restricted discrepancy statistics \dcphr{D} and \dcpir{D}, which are found by averaging over the correlations only over bonds that meet the restriction 
$\left|\langle s_is_j\rangle_{\sigma_1}+\langle s_is_j\rangle_{\sigma_2}\right|<1$ and $\left|\left[s_is_j\right]_{\eta_1}+\left[s_is_j\right]_{\eta_2}\right|<1$, respectively.
The ratio of the global discrepancy to the restricted discrepancy should be given by the domain wall density, if there is not a singular contribution from nearly-rigid bonds to the unrestricted discrepancy statistic. 
We find that the
discrepancy statistics \dcpr{D} vary with the amount of samples $n_{\eta,\sigma}$. More specifically,
the measured discrepancy statistics \dcp{D} and \dcpr{D} are found to be well fit by a correction proportional to $1/n_{\eta,\sigma}$, consistent with simple random sampling error of the $\langle s_i
s_j\rangle$ and $\left[s_is_j\right]$ correlation functions via Bernoulli statistics. We therefore estimate $\mathcal D(n_{\sigma,\eta}=\infty)$ and
$\mathcal{D}^r(n_{\sigma,\eta}=\infty)$ for each $L$ and $\ell$ by fitting $\mathcal D(\ell)$ and $\mathcal D^r(\ell)$ for each ensemble to $D(\ell,n)=D(\ell,n=\infty)+1/n$, using data for $n_{\sigma,\eta}=250,500,1000$. 

\begin{figure}
	\centering
	\includegraphics[width=\columnwidth]{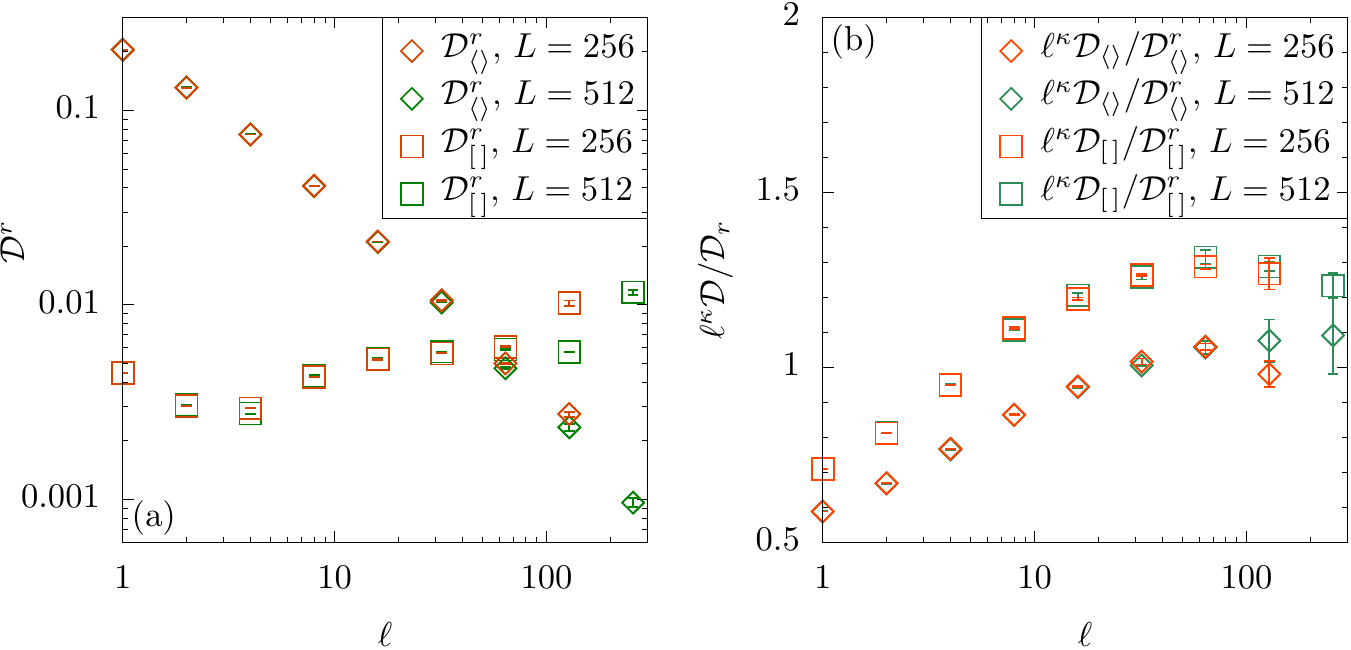}
	\caption{(Color online).
		Plots based on the discrepancy statistics $\mathcal D$ and $\mathcal D_r$. These statistics characterize the variation in
        the nearest-neighbor correlations of the 2DISG, where the correlation differences are averaged over either all bonds or the restricted set of ``non-rigid'' bonds, respectively.   (a) Plots of the noise-history-averaged \dcphr{D} and \dcpir{D}, extrapolated to $n=\infty$. The discrepancy statistic \dcphr{D} measuring differences in correlations between two initial states, the correlations found by averaging over noise histories for each state, decays with coarsening scale $\ell$.
        The discrepancy statistic \dcpir{D} computed by comparing two noise histories, with correlations found by averaging over initial configurations for each noise history, appears to remain finite 
        at long times. (b) The ratios ${\mathcal D}_{\langle\rangle}/{\mathcal D}_{\langle\rangle}^r$ and ${\mathcal D}_{[\,]}/{\mathcal D}_{[\,]}^r$ appear to scale 
		as $\ell^{-\kappa}$, since the data $\ell^{\kappa}\mathcal D/\mathcal D_r$ converge for large patch scales $\ell$, supporting the supposition that the discrepancy in correlations is dominated by the fractal domain walls, for both noise-averaged and initial-configuration-averaged correlations.
	}
	\label{fig:discrepancy}
\end{figure}

Data
for the dependence of the restricted discrepancy statistics \dcpr{D} on coarsening scale $\ell$ are plotted in \figref{fig:discrepancy}(a), for $L=256,\,512$, using this extrapolation $n\rightarrow\infty$.
The noise-history-averaged restricted discrepancy statistics 
\dcphr{D} decay as patch scales $\ell$ increase, implying that the noise-averaged local correlations are gradually erased and become independent of the initial state $\sigma$. In contrast, the 
restricted discrepancy statistic \dcpir{D} changes slowly at small scales (compared to the statistic \dcphr{D}) and then
approaches a constant \dcpir{D}$\approx 0.2^2$, indicating a typical scale-independent difference between noise histories of size $\approx0.2$ in the domain-wall correlations. These extrapolations are consistent with the vanishing variations visible in \figref{fig:nbCorI_v} and the remnant variations in the correlations visible in for the two noise histories shown in \figref{fig:nbCorP_v}. To confirm that the dominant contribution to the unrestricted statistic is confined to the domain walls, we plot the scaled ratios
$\ell^\kappa{\mathcal D}_{\langle\rangle}/{\mathcal D}_{\langle\rangle}^r$ and $\ell^\kappa{\mathcal D}_{[\,]}/{\mathcal D}_{[\,]}^r$ in \figref{fig:discrepancy}(b), using the value
$\kappa=2-d_f$, with
$d_f=1.27$ \cite{Middleton2001,Hartmann2002} for the dimension of the domain walls that separate the rigid domains. The convergence of this ratio to a nearly constant value is consistent with the restricted discrepancy, associated with the domain walls, being the primary contributor to the unrestricted statistics, with the rigid domains separated by domain walls that are seen statistically by averaging over multiple initial configurations or noise. We conclude that while a selected noise history does not fix the configuration, as there is variability given by the initial state, the initial-state-averaged correlations in a given domain wall are set by the noise history, at each scale of coarsening.

In this work so far, we have studied memory of initial conditions and noise history by working at zero temperature and using a heuristic coarsening. We have argued that this
heuristic approach should replicate low-temperature behavior in the spin-glass phase or at shorter length scales, for scale less than the correlation length, in the case of a
zero-temperature transition, as holds for the 2DISG. We include here initial work that supports this correspondence. We have carried out low-temperature single-spin flip Glauber
dynamics for the 2DISG, using updates of the spins selected in random order, for up to $5^9$ sweeps in lattices of linear size $L=150$. For each temperature, we compute a time-dependent
correlation length defined by the inverse of the correlation function of the overlap of the spin configurations with the ground-state configuration, as used in Ref.\ \cite{Kisker1996}. We can
then plot the aging and heritability as a function of correlation length $\xi$, as in the zero-temperature studies, though, due to the much slower dynamics, we reach 
only about one-half of a decade in range of length scale. Our results are plotted in \figref{fig:q_T}. We show straight lines that give the power-law decays seen in our 
zero-temperature work as a comparison. While we do not have a long range of length scales to compare, these results indicate that the zero-temperature persistence of 
configurations and the heritability of configurations decays with coarsening scale in a fashion not inconsistent with finite-temperature results.

\begin{figure}
	\centering
	\includegraphics[width=\columnwidth]{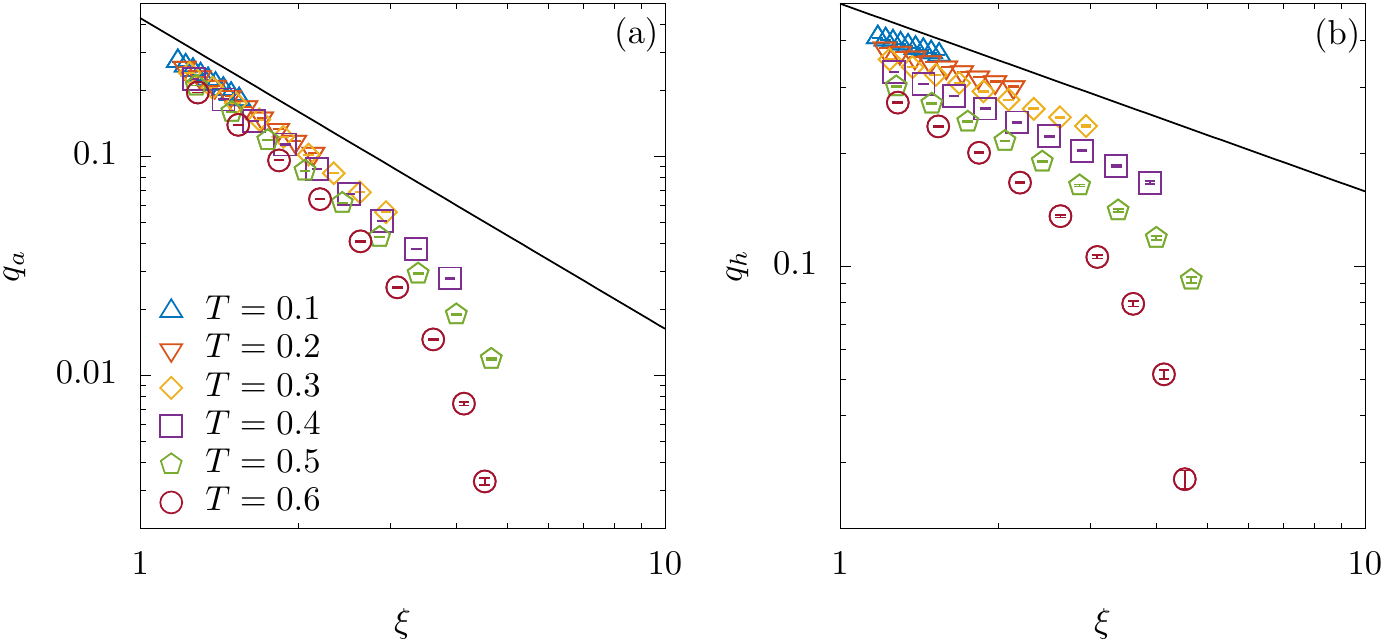}
    \caption{(Color online). Plots of the sample-averaged spin overlap statistics $q_a$ and $q_h$ in (a) the aging study and (b) the heritability study at finite temperatures as
    spin glasses relax under the Glauber dynamics. The waiting times are $t_w=5^n$ ($n=1,\ldots,9$). For each waiting time, the spin overlaps ($q_a$ and $q_h$) and the correlation
    length $\xi$ are calculated. The time-dependent correlation length $\xi$, indicating the coarsening of spin configurations, is determined by comparing spin configurations with the ground
    state. The spin overlaps $q_a$ and $q_h$ are compared with power law decays at low temperatures, $q_a\sim\xi^{-1.42}$ and $q_h\sim\xi^{-0.5}$, using the values from the scale-dependence seen in
    patchwork dynamics at zero temperature \cite{Yang2017}. At the higher temperatures, the divergence from the power law behavior, due to the finite correlation length at any finite temperature $T$, is clearly seen, while lower temperature results are not inconsistent with the values from patchwork dynamics.
	}
	\label{fig:q_T}
\end{figure} 
 
In conclusion, we have used patchwork dynamics to investigate the relative importance of initial states (``nature"), noise histories (``nurture") and intrinsic disorder
(``environment") in determining the long-time nonequilibrium states of a glassy spin model. A heritability study shows that the similarity of identical systems decays as a power law
with the length scale. A heritability exponent is estimated as $\theta_h\simeq0.5$. The heritability exponent for the 2DISG is close to the persistence exponent $\theta_p$ that 
describes the power low decay of non-flipped spins in patchwork dynamics. The reference \cite{Ye2013} finds that the heritability and persistence exponents are 
very close for the 2D Ising model, similar to our result here. We study the effect of nature, 
nurture and environment on the average nearest-neighbor correlation configuration. Under patchwork dynamics, the configuration exhibits coarsening, with rigid domains (independent of noise or initial configuration) separated by domain walls determined by 
the intrinsic disorder of the 2DISG sample and the coarsening scale $\ell$. A restricted discrepancy statistic is defined to quantify the influence of initial states and noise histories. Using this statistic, we
find that the initial state information is erased by averaging over noise histories. In contrast, distinct noise histories give rise to distinct correlations, at all scales, when the correlations are found by averaging over initial states. This dependence of correlations on noise history at zero temperature, and not initial states, may be linked to the convergence of all microscopic configurations under a single noise history, seen at temperatures above a certain value by Chanal and Krauth \cite{ChanalKrauth2008}. We also provide initial direct numerical evidence that patchwork dynamics provides an efficient heuristic algorithm to mimic the long time scales associated with nonequilibrium behavior in the 2DISG, albeit we can display this over only half of a decade in length scale, due to the glassiness of low-temperature equilibration.

This work was supported in part by National Science Foundation Grant No. DMR-1410937. Simulations were performed primarily on the Syracuse University campus 
OrangeGrid, a computing resource supported by NSF Award No. ACI-1341006.


\end{document}